\input amstex
\documentstyle{amsppt}
\loadmsbm
\def\today{\ifcase\month\or  January\or February\or March\or April\or
May\or June\or July\or August\or September\or October\or November\or
December\fi \space\number\day, \number\year}
\def\cmp{Comm.\ Math.\ Phys.}
\def\prb{Phys.\ Rev.\ B}
\def\ap{1}
\def\arz{2}
\def\asy{3}
\def\ass{4}
\def\bellissard{5}
\def\birman{6}
\def\bw{7}
\def\bmm{8}
\def\carey{9}
\def\ct{10}
\def\connes{11}
\def\cuntz{12}
\def\cfks{13}
\def\dn{14}
\def\efros{15}
\def\fedosov{16}
\def\fradkin{17}
\def\f{18}
\def\hormander{19}
\def\ka{20}
\def\kato{21}
\def\kg{22}
\def\ks{23}
\def\kohmoto{24}
\def\kunz{25}
\def\laughlin{26}
\def\matsui{27}
\def\nb{28}
\def\niu{29}
\def\nt{30}
\def\ntw{31}
\def\pg{32}
\def\russo{33}
\def\seiler{34}
\def\sw{35}
\def\trace{36}
\def\semigroup{37}
\def\stone{38}
\def\streda{39}
\def\tknn{40}
\def\thouless{41}
\def\wen{42}
\def\wigner{43}
\def\wilczek{44}
\def\xia{45}
\def\zak{46}

\NoRunningHeads
\hsize=36pc
\vsize=54pc
\topmatter
\title
{Charge Deficiency, Charge Transport and  Comparison of Dimensions}
\endtitle
\TagsOnRight
\vfill
\break
\author
Joseph E. Avron${}^1$,
Ruedi Seiler${}^2$ and
Barry Simon${}^3$\endauthor
\affil ${}^1$  Department of Physics, Technion, Israel Institute of
Technology\\
${}^2$ Fachbereich Mathematik, Technische Universit\"at Berlin\\
${}^3$ Division of Physics Mathematics and Astronomy,
Caltech\endaffil
\address ${}^1$ Technion City, Haifa, 32000, Israel.\newline
${}^2$ Strasse des 17. Juni, W-1000 Berlin 12, Germany.\newline
${}^3$ Pasadena, Ca 91125, USA \endaddress
\email ${}^1$ phr97ya\@technion.ac.il,
${}^2$ seiler\@math.tu-berlin.de\endemail
\date \today\enddate
\thanks
Part of this work was written while the authors enjoyed the
hospitality of the Landau Center at the Hebrew University.  The
work is supported by BSF, DFG, GIF, NSF, the fund for the
Promotion of Research at the Technion and the Technion VPR--
Steigman research fund. One of the authors (RS) should like to
acknowledge the
hospitality of  the Mittag-Leffler  Institute.
\endthanks
\abstract
{We study the relative index of two orthogonal
infinite dimensional projections which, in the finite dimensional
case, is the difference in their dimensions. We relate  the relative
index to the Fredholm index of  appropriate operators,  discuss
its basic properties,  and obtain various formulas for it.  We apply
the relative index to counting the change in the number of
electrons below the Fermi energy of certain quantum systems and
interpret it as the charge deficiency.  We study the  relation of
the charge deficiency with the notion of adiabatic charge transport
that arises from the consideration of the adiabatic curvature.  It is
shown that, under a certain covariance,  (homogeneity),  condition
the two are related.  The relative index is related to Bellissard's
theory of
the Integer Hall effect. For Landau Hamiltonians the relative index
is
computed explicitly for all Landau levels.}
\endabstract
\endtopmatter
\document
\heading {\bf 1. Introduction} \\ \endheading  \parindent=0pt
\parskip=6pt
An interesting observation that emerged in the last
decade is that charge transport in quantum mechanics, in the
absence  of dissipation, often lends itself to geometric
interpretation. A good part,  but not all, of this research has
been motivated by, and applied to, the integer and fractional Hall
effect
\cite{\arz ,\bmm ,\connes ,\fradkin ,\ka ,\laughlin ,\pg
,\sw ,\seiler ,\stone ,\wilczek}.  \par The framework that will concern
us
here is that of (non-relativistic) quantum mechanics.  Within this
framework  common models of the integer Hall effect are Schr\"odinger
operators  associated
with non interacting electrons in  the plane, with (constant)
magnetic field
perpendicular to the plane and  random
(or periodic) potential.   The Hall conductance has been
related to a Fredholm Index by Bellissard \cite{\bellissard }, and to
a Chern
number by Thouless, Kohmoto, Nightingale and den-Nijs \cite{\tknn
}.
The Fractional Hall effect is associated with electron-electron
interaction
and this goes beyond what we  do here. \par
Quantum  field theory is another framework where transport
properties and
geometry are related.  The focal point here has been the  Fractional
Hall
effect and  the associated Chern-Simons field  theories \cite{\bw ,
\bmm ,\f ,\laughlin ,\wen,\wilczek} . We shall not address these
issues.

\par  The Chern number approach to quantum transport has been
extended  to a large class of quantum mechanical systems, including
models of the integer Hall effect \cite{\fradkin ,\kohmoto ,\kunz ,
\niu ,\nt ,\ntw ,\thouless }, to  models
with electron-electron interactions \cite{\asy ,\ks ,\nt } and to
other
systems that bear only little resemblance to the Integer Hall effect
\cite{\bmm ,\dn ,\niu ,\sw ,\stone}. The Index approach has not
been as
popular, and  has not been substantially extended beyond the one
electron
setting  considered by  Bellissard for the integer Hall effect
\cite{\bellissard ,\connes ,\nb ,\xia }. \par
We have  two main purposes in this work.  The first is to  develop
the Index approach  from the physical point of view of  ``charge
deficiency":  Consider a  quantum system of (non-interacting) electrons
where  the Fermi energy is in a gap. We allow an infinitely large number
of
electrons  below the Fermi energy. Now consider taking this system through
a  cycle, so that at the end of the cycle the Schr\"odinger operator is
unitarily related to the one at the outset. The examples we shall focus
on
here are where  the initial and final systems are related by a {\it singular}
gauge transformation corresponding to  piercing the system with  an
infinitesimally thin  flux tube, carrying one unit of  quantum flux.
Because of the unitary  equivalence, at the end of the cycle we can put
the
Fermi energy  in the same  gap as at the outset, and can ask for the
difference in the number of  electrons below the Fermi  energy. This
deficiency  of charge counts the charge  transported in or out of the
system
as a result of  the additional flux quantum.  In interesting cases this
difference is $\infty-\infty$. For non-interacting electrons, such a
difference is   the difference in dimensions of a  pair of two infinite
dimensional  Hilbert space projections.  This is the relative  index.
It turns
out to  be related to  an index of an appropriate Fredholm operator. In
particular, it is an integer. (The charge deficiency introduced here is
reminiscent of a charge that enters in computing  the vacuum polarization
in
Fock space. See \cite {\matsui }.)\par  The identification of  charge
deficiency with an index implies integral charge transport. This holds
for a
wide class of  two dimensional quantum system, including the conventional
models of the integer  Hall effect mentioned above.  But it also holds
for
also  more general  models whose  geometries and background  potentials
may be far removed  from the Integer Hall effect. \par The theory described
below appears to be restricted, at the moment at least, to non-interacting
electrons.  This  is consistent with the common wisdom because
electron-electron interaction will, in general, lead to fractional transport.
\par  Our  second purpose is to examine the relation of the charge deficiency
(associated with an index) and  the  notion of charge transport that
arises in theories  of  linear and adiabatic response.  The latter is
associated  with Kubo's formulae,  Chern numbers and adiabatic curvatures.
These  two notions are distinct in general.  They turn out to be related
for
homogeneous systems. These are the  kind of systems relevant to the Integer
Hall effect.

This relation between charge deficiency and charge transport is
reminiscent of known identities in related contexts: \v Streda's
formula (which is relating that the Hall conductance  with a  gap label)
\cite{\streda
} and certain Ward identities in  Chern-Simons  fields theories
giving rise to
relations between transport coefficients in  linear  response theory
\cite{\f ,\wen }. \vskip 0.3in
\proclaim{Acknowledgment}
We are grateful to S.~Agmon, E.\ Akkermans,  J.\ Bellissard, S.\ Borac,
J.\ Fr\"ohlich, I.\ Kaplansky, M.\ Klein, A.\ Pnueli and U.~Sivan for
useful
discussions and comments. \endproclaim \vskip  0.3in

\heading {\bf 2. Comparing Dimensions} \endheading
\vskip 0.3in
In this section we describe various formulas for comparing
dimensions of two orthogonal projections, $P$ and $Q$. The index
for two projectors of finite rank is just the difference of their
dimensions.
$${Index}\,  (P,Q) \equiv {dim}\, P - {dim}\, Q
 = {Tr} \, (P-Q)\tag2.1$$
A possible and, as we shall see, natural generalization of (2.1) to
the infinite dimensional case is:
\proclaim{Definition (2.1)}
Let $P$ and $Q$ be orthogonal projections so that  $P-Q$ is compact,
then
$${Index}\,  (P,Q) \equiv {dim} \big( Ker\, (P-Q-1)\big)- {dim}
\big(Ker\, (Q-P-1)\big) .\tag2.2$$  \endproclaim
This Index is a well defined  finite integer since
${dim} \big( Ker(P-Q\pm 1)\big)$ are both finite by the
compactness of $P-Q$. (One could take a broader perspective and define
the left
hand side of 2.2 by the right hand side whenever the latter makes sense).
Before we discuss in what  sense 2.2 is a
generalization of 2.1 we note that the relative index indeed has
some of the natural properties of an object that compares  dimensions
of
two projections:  $$\eqalign{{Index}\, (P,Q) =
- {Index}\,  (Q, P) &
 =- {Index}\, (P_{\bot} , Q_{\bot})= {Index}\,  (UPU^{-
1},UQU^{-1}),\cr
P_{\bot} &\equiv 1 - P{,} \quad Q_{\bot} \equiv 1 - Q,}
\tag2.3$$
for any linear and invertible map $U$.
The basic formulas for computing the relative Index is:
\proclaim{Proposition (2.2)}
Suppose that $\allowmathbreak (P-Q)^{2n+1}$ is trace class for a
natural number $n$, then
$${Index}\,  (P,Q) = {Tr}\,  (P-Q)^{2n+1}.\tag2.4$$
\endproclaim
It follows that the right hand side of 2.4  is independent of $n$
for $n$ large enough,  and that it
reduces to 2.1 in the finite dimensional case.  We shall return to
the proof of this proposition shortly.\par  To see where (2.4) comes,
we
start by  noting an algebraic identity for any pair of projections $P$
and $Q$:
$$(P-Q)^2 P = P-PQP = PQ_{\bot} P =P(P-Q)^2.\tag2.5$$
In particular this says that $(P-Q)^2$ commutes with $P$ and $Q$.
This leads to:
\proclaim{Proposition (2.3)} Let $n$ be a nonnegative integer
so that $(P-Q)^{2n+1}$ is trace class, then:
$${Tr}\, (P-Q)^{2n+3} = {Tr}\, (P-Q)^{2n+1}.\tag2.6$$
\endproclaim
\demo{Proof}
Subtracting the two equations below from each other
$$\eqalign{(P-Q)^{2n+2}P &= (P-Q)^{2n}(P-PQP) \cr
(P-Q)^{2n+2}Q &=(P-Q)^{2n}(Q-QPQ),}\tag2.7
$$
gives
$$(P-Q)^{2n+3} =(P-Q)^{2n+1} - (P-Q)^{2n}[PQ, QP].\tag2.8$$
Since: $$[PQ,QP] = \Big[ PQ, [Q,P]\Big]
=\Big[ PQ, [Q, P-Q]\Big],\tag2.9$$
we get, due to equation (2.5), the identity:
$$(P-Q)^{2n+3} = (P-Q)^{2n+1}-[PQ,B],\quad
 B \equiv \left[ Q, (P-Q)^{2n+1}\right].\mathstrut \tag2.10
$$
$PQ$ is bounded  and $B$ is trace class, so ${Tr}\, [PQ,B] = 0$. Tracing
(2.10)
gives (2.6).
\qed
\enddemo
In the applications we never  go beyond the trace class situation
discussed  above, in fact the case $n=1$ covers all the cases we shall
consider. \par
\demo{Proof of Proposition 2.2} (2.6) implies that
$Tr\,\big((P-Q)^{2m +1}\big)$ is independent of $m$ for
$m\ge 0$.   As m goes to
infinity, this trace converges to $Index\, (P,Q)$ since \hbox{$-1 \le
P-Q \le
1$}. Thus (2.4) is proven.\qed
\enddemo
In future work we'll examine this result further providing several
other proofs which illuminate it.

In the applications we consider projectors  $P$ and $Q$ on
subspaces with energies below some fixed Fermi energy.  ${Index}\,  (P,Q)$
then counts the difference in the number of electrons,
which we identify with the charge
deficiency.  Physical considerations, that we shall  describe in
the following sections, motivate considering  $P$ and $Q$ which are
related by a  unitary $U$: $$Q =
UPU^*. \tag2.11$$
In the finite dimensional case $P$
and
$Q$  are related by a unitary if and only if their dimensions coincide.
In the
infinite dimensional case of a separable Hilbert space with
${dim} P = {dim} P_{\bot} = {dim}  Q = {dim}
Q_{\bot} = \infty$ such a $U$ always exists, and  does not force
${Index}\, (P,Q) =0$.

In the case that P and Q are related by a unitary map the index of
the pair can be related to a Fredholm index of one single operator:
\proclaim{Proposition(2.4)}
Let $Q = UPU^*$, $P$ an orthogonal projections and $U$ unitary and
suppose that $(P-Q)^{2n+1}$ is trace class. Then,
${Tr}\,  (P-PQP)^{n+1}$ and  $ {Tr}\, (Q-QPQ)^{n+1}$ are trace class;
$PUP$ is a Fredholm operator in range P and
$$\eqalign{ &{Index}\, (P,Q)= {Tr}\,
([P,U]U^*)^{2n+1} = {Tr}\, (P-PQP)^{n+1} - {Tr}\, (Q-QPQ)^{n+1} \cr
&= -\Big( dim Ker (U|Range P) -dim Ker (U^*|Range
P)\Big)\equiv -{Index}\, (PUP).}\tag2.12$$  \endproclaim
\demo{Proof}
The first identity is a rewrite of (2.4) upon noting that
$$P-Q =[P,U]U^*.\tag2.13$$
The second identity follows from (2.5) which gives:
$$\align (P-PQP)^{n+1} &= ((P-Q)^2P)^{n+1}    = (P-Q)^{2n+2}P \\
         (Q-QPQ)^{n+1} &= ((P-Q)^2Q)^{n+1}
= (P-Q)^{2n+2}Q,\tag2.14\endalign$$
(proving our trace class assertion), subtracting and tracing using
(2.4) and
(2.6) gives the  second identity.  To get the third identity note that:
$$\align P-PQP &= P-PUPU^*P\\
Q-QPQ &=U(P-PU^*PUP)U^*,\tag2.15
\endalign$$
using the unitary invariance of the trace we see
that the third term in (2.12) can be written as:
$${Tr}\, (P-PUPU^*P)^{n+1}-{Tr}\,  (P-PU^*PUP)^{n+1}.\tag2.16$$
Since both terms are
finite the operators $(PUP)$ and $(PU^*P)$ are inverses of each
other in range P up to compacts. A formula of Fedosov
\cite{ \hormander ,\fedosov } then says that under such
circumstances
(2.16) is a formula for ${Index}\,  (PU^*P)$ respectively
$-{Index}\, (PUP)$.
\qed\enddemo
We can now use the relation
${Index}\,  (P,Q) =-{Index}\,  (PUP)$, to transfer known facts
about the Fredholm Index to the relative index, and vice versa.
\proclaim{Proposition (2.5)} Let $P,Q,R$ be orthogonal projections, which
differ by  compacts. Then
$${Index}\, (P, R) = {Index}\, (P,Q) +{Index}\, (Q,R).\tag2.17$$
\endproclaim
This identity is, of course, trivial in the situation where
 $P,Q,R$ differ by trace class operators. When interpreted as
charge deficiency, it is a statement of charge (or particle)
conservation.
\demo{Proof}  For simplicity we suppose
that $P,Q$ and $R$ are unitarily related.  Elsewhere we  shall give a
proof of the general case.
\par
Eq. 2.17 equivalent to:
$$ {Index}\, (P(U_2U_1)P) = {Index}\, (PU_1P)+
{Index}\, (QU_2Q).\tag2.18$$
Now we rewrite all expressions in terms of Q and the necessary unitaries:
$$\align
{Index}\,(PU_2U_1P) &= {Index}\,(U_1^{-1}QU_1U_2QU_1)\\
                    &= {Index}\,(QU_1U_2Q)\\
{Index}\,(PU_1P)    &=  {Index}\,(U_1^{-1}QU_1QU_1)\\
                    &= {Index}\,(QU_1Q)
\endalign$$
Hence it remains to show
$$ {Index}\,(QU_1U_2Q) = {Index}\,(QU_1Q) + {Index}\,(QU_2Q)\tag2.19$$
The left hand side can be replaced by ${Index}\,(QU_1QU_2Q)$ because
the difference of the corresponding operators is compact,
$$ QU_1QU_2Q - QU_1U_2Q = Q[U_1,Q]U_1^{-1}U_1U_2Q.$$
This follows from the compactness of $[U_1,Q]U_1^{-1} $
and the fact that all the remaining terms are bounded. By a basic
result of stability theory for indices \cite{\kg } the index is invariant
under perturbations by compacts. Furthermore by the product formula for
Fredholm indices one gets
$$ {Index}\,(QU_1QU_2Q) = {Index}(QU_1Q) + {Index}\, (QU_2Q), \tag2.20$$
This proves the proposition. \qed\enddemo
Related questions are addressed in \cite{\carey ,\cuntz ,\efros }.
\vskip 0.3in

\heading {\bf 3. Gauge Transformations and Computations with
Integral Kernels}
\endheading \vskip 0.3in
In this section we introduce additional structure into the
general operator theoretic framework of the previous section,
which will  accompany us throughout.  It is motivated by the
applications
we have in mind, and  involves conditions on the kind of projections we
consider and the unitaries  that relate them. In particular, the
unitary that relates the orthogonal projections $P$ and $Q$ will be
associated  with a (singular) gauge transformation which corresponds to
piercing the  quantum system with a flux tube carrying an integral number
of flux quanta.  That is, $U$ is  a unitary  multiplication operator whose
winding is the  number of flux quanta carried by the flux tube. (More
precise
conditions will  be stated shortly). This naturally forces us into considering
two dimensional quantum systems. Furthermore,  it turns out, that for
${Index}\, (P,Q) \neq 0$ the orthogonal projection $P$ has to be infinite
dimensional and time reversal invariance must be broken.  \par We describe
this additional structure under \proclaim{ Hypothesis (3.1)} \par {(a)}
The Hilbert space is $L^2(\Omega)$ where $\Omega \subseteqq {\Bbb
R}^2$ is a
two dimensional domain in $\Bbb R^2$ with smooth (possibly
empty) boundary  $\partial \Omega$. In particular, the orthogonal
projections $P$ and $Q$ of the previous  section are projections in $L^2
(\Omega)$. \newline {(b)} The projection $P$
has integral kernel $p(x,y)$, $x,\, y \in \Omega$,
which is jointly continuous in $x$ and $y$ and decays away from
the diagonal, so that:
$$|p(x,y)| \leq \frac{C}{1 + \big(dis(x,y)\big)^\eta}\tag3.1$$
with $\eta > 2$ and $dis(x,y)$ is the distance between
$x$ and $y$.
\newline {(c)} $U$ is a multiplication operator on
$L^2(\Omega)$
by a complex valued function $u(x)$, with $|u(x)| = 1$, and $u(x)$ is
differentiable away from a single point  which we take to be $x=0$.
The derivative is
$\Cal O (\frac{1}{|x|})$. More precisely, we assume that there are
constants $C_1$ and $C_2$ such that:
 $$|u(x+y) - u(y)| \leq C_1 \,\frac{|x|}{|y|}\tag3.2$$
for $|x| \le C_2\, |y|$. The winding
number of $U$ about the singularity is denoted by $ N(U)$.  This is
the number of magnetic flux quanta carried by the flux tube
associated with $U$.
\endproclaim
\proclaim{Example (3.2)}
Let $\Omega = \Bbb R^2$, and let $z=x+i\,y.$
$$ u_\alpha (z) = \left\{
                        \aligned
                                \frac{z^\alpha}{|z|^\alpha} ,
                                \qquad &z \in \Bbb R^2 /[0,\infty]\\ 1,\qquad
&z\in [0,\infty)
                        \endaligned \right. \tag3.3
$$
are unitaries which, for integer $\alpha$,  are smooth away from
the origin and have winding number $\alpha$.  Such unitaries are
associated with an infinitesimally thin flux tubes through the origin
carrying
$\alpha$
units of quantum flux.  In particular, for $\alpha =1$ condition c
above holds with $C_1 = C_2^{-1} = 2$. This follows from the elementary
inequality  $| u_1(z) - u_1(z')| \leq |z - z'| \max
(\frac{1}{|z|},\frac{1}{|z'|})$, which implies (3.2).
On the other hand, if $\alpha \not\in \Bbb Z$, condition c clearly
fails near the positive real axis.
\endproclaim
The fact that $U$ is a gauge transformation distinguishes coordinate
space, and in the rest of this section the integral kernel of $P-Q$
will play a
role.  In particular, we'd like to know that an object like
 ${Tr} \, (P-Q)^3$ can be computed from the integral kernel of $P-Q$
by integrating  on the diagonal. This somewhat technical issue is
guaranteed by the following preparatory
result: \proclaim{Proposition (3.3)} Let $K$ be trace class with
integral kernel $K(x,y),$\ $x,y \in {\Bbb R} ^n$, which is jointly
continuous in
$x$ and $y$ away from a finite set of points $(x_i, y_i)$ so that
$K(x,x)\in L^1$ in neighborhoods of these points, then: $${Tr}\,  K
= \int\limits_{\Bbb R^n}
K(x,x) \,dx\tag3.4$$ \endproclaim \demo{Sketch of proof} Let
$E_\epsilon$, $F_\epsilon$, $G_\epsilon$  be the
characteristic functions of the union of $\epsilon$-balls about the
singular points, the exterior of a $1/\epsilon$ ball and the
complement of these two sets.  Then
$$Tr\, (K) =  Tr\, (E_\epsilon K) + Tr\, (F_\epsilon K) + Tr\,
(G_\epsilon
KG_\epsilon)\eqno(3. 5)$$ where we used cyclicity of the trace to
get the last term.  Since $E_\epsilon$ and $F_\epsilon$ converge
strongly to
zero as $\epsilon$ goes to 0, $E_\epsilon K$ and $F_\epsilon K$ go
to zero in trace norm (as can be seen by writing $K$ as a finite rank
plus
small trace norm), and since a result in \cite{\trace } says that $Tr\,
(G_\epsilon KG_\epsilon )$ is the integral over $G_\epsilon$ of
$K(x,x)$ the result follows by taking the limit using the fact that
$K(x,x)$ is $L^1$.  This proves proposition (3.3).
\qed\enddemo
Proposition (3.3) could be replaced by the following statement
which is is a kind of a Lebesgue integral version of proposition 3.3
\cite{\birman }.  Its application to the concrete cases we have in
mind requires however somewhat more care.
\proclaim{Remark (3.4)}
Let K be trace class on $L^2(\Bbb R^n)$. Then,  its integral kernel
$K(x,y)$ may be chosen so that the function $L(x,y) \equiv
K(x,x+y)$ is a continuous function of $y$ with values in $L^1(\Bbb
R^n)$.
Furthermore if $l(y)\equiv \int L(x,y)\, dx$ then $Tr\,K = l(0)$.
\endproclaim
Our first application is the following result that guarantees that
${Index}\,  (P-Q)=0$  if $P-Q$ is trace class:
\proclaim{Proposition (3.5)}
Suppose $P-Q$ is trace class with $Q=UPU^{-1}$, $U$ and $P$
satisfying hypothesis (3.1). Then ${Index}\, (P,Q) = {Tr}\, (P-Q) = 0$.
\endproclaim
\demo{Proof}
The integral kernel of $P-Q$ is:
$$(P-Q)(x,y) = p(x,y)\Big(1-\frac{u(x)}{u(y)}\Big)\tag3.6$$
By proposition (3.3), ${Tr} \, (P-Q)$ is the integral
of (3.6) on the diagonal with $x=y$. But the kernel of
$(P-Q)$ vanishes on the diagonal. Hence the trace is zero.
\qed\enddemo
This means the trace class situation is like the finite dimensional
case, i.e. unitary equivalence of $P$ and $Q$ implies
equality of dimensions in the generalized sense.
In particular, for ${Index}\, (P,Q) \neq 0$, $(P-Q)$ must not be trace
class, so $dim\, P =dim \, Q =\infty$.

The following proposition is central.
 \proclaim{Proposition (3.6)} Under hypothesis
(3.1) \, $(P-Q)^p$ is trace class for $p>2$. In particular \hbox{${Tr}\,
(P-Q)^3$} is an integer and
$$-{Index}\, (PUP) =
\int\limits_{\Omega^3} \, dx \, dy \, dz \, p(x,y)p(y,z)p(z,x)
\left(1-\frac{u(x)}{u(y)}\right)\left(1-
\frac{u(y)}{u(z)}\right)\left(1-\frac{u(z)}{u(x)}\right).
\tag3.7$$ \endproclaim
\proclaim{Remarks (3.7)} 1. In the case where $p(x,y)$ is
$C^\infty_0$ the proposition is in \cite{\connes }. \par
2. The index is real, of course. Under complex conjugation the first
triple product in 3.7 becomes  $p(y,x)p(z,y)p(x,z)$, since, by self
adjointness $\bar p (x,y) =p(y,x)$.  The second triple product transforms
to
$\big(1-\frac{u(y)}{u(x)}\big)\big(1-\frac{u(z)}{u(y)}\big)\big(1-
\frac{u(x)}{u(z)}\big)$ by the unimodularity of $u(x)$.
This reduces to the original integrand upon interchanging $x$ and
$z$.

3. If we were to consider, for example, $\Bbb R^3$, then the
integrand in 3.7, under hypothesis 3.1, would lack decay in the
direction of the magnetic flux tube, and 3.7 would  become
meaningless,
in  general.

4.  Flux tubes that carry fractional fluxes are associated with
unitaries of example 3.2 with $\alpha \not\in \Bbb Z$. For such
$U$'s, the integrand in 3.7 lacks decay along the cut, and the integral
is
divergent in
general.

5.  This proposition also tells us that, as far as section 2 is
concerned, $n=1$ is all we have to consider. \endproclaim

\demo{Proof}
By hypothesis (3.1) $P-Q$ is an integral operator with kernel
$p(x,y)\left(1 - \frac{u(x)}{u(y)}\right)$. To prove that $(P-Q)^p \,
,p > 2,$ is trace class it is enough to show that the function
$$g(y) \equiv \int \vert p(x + y,y) \left(1 - \frac{u(x + y)}
{u(y)}\right)\vert^q\,dx \in L^{p-1}(\Bbb R^2),\qquad 1/p + 1/q =
1,\tag 3.8$$ because of Russo's theorem \cite{\russo }. To prove (3.8)
notice that close to the diagonal $x=0$ the second term of the integrand
in
(3.8)  is small, off the diagonal it  is
the first one which is small. To put this in analytic form we note
firstly that it is enough to  prove (3.8)
in the following situation:  Replace in (3.8)
$ p(x + y,y) \left(1 - \frac{u(x + y)}{u(y)}\right) $
by  the function
$$f(x,y) \equiv \frac {1}{1 + \vert x \vert^\eta}
Min \{C_2,\frac {\vert x \vert}{\vert y \vert}\} \tag3.9$$
where $C_2$ is the constant introduced in hypothesis (3.1);
i.e. it is enough to prove
$$F(y)\equiv \int \Big(f(x,y)\Big)^q \, dx \in L^{p-1}(\Bbb R),
\tag3.10$$
because, by construction,
$ \vert p(x + y,y) \left(1 - \frac{u(x + y)}{u(x)}\right)\vert $
is pointwise dominated by a constant multiple of $f(x,y)$.
Secondly we show that $F$ is uniformly bounded in y.This follows
from the y independent bound on $f(x,y)$,
$$\left(f(x,y)\right)^q \leq const \left(\frac{1}
{1 + \vert x \vert^\eta}\right)^q \,\tag3.11$$
together with
$\eta q - 2 > 0 $ ( use $\eta > 2$ and $ q > 1$ ).
Hence the right hand side of 3.10 is integrable. Thirdly we analyze
the behavior of $F$ for large $y$. To do that we
split the defining integral into two pieces and prove that each term
is in $L^{p-1}(\Bbb R)$. The first term is defined by
$$F_1(y)\equiv \int _{I(y)}\Big(f(x,y)\Big)^q \,dx ,\tag3.12$$
where $I(y) \equiv \{x \vert \,\,{\vert x \vert }\leq {C_2 }\vert y
\vert\}$ denotes the domain close to the diagonal $x=0$.
By construction it satisfies the estimate
$$F_1(y)\leq \frac {1}{\vert y \vert^q} \int _{I(y)}\frac{\vert
x\vert^q} {(1 + \vert x\vert^\eta )^q}\, dx.\tag3.13$$
Cutting out the unit ball $B$ \,in $I(y)$  we get the inequality
$$F_1(y)\leq \frac{\pi}{{\vert y \vert}^q } + \frac {1}{{\vert y
\vert}^q}
\int _{I(y)\setminus B}\frac{{\vert x\vert}^q }
{(1 + \vert x\vert^\eta )^q}\, dx. \tag3.14$$
The second term is bounded  up to a constant $2\pi$ by
$$\frac{1}{\vert y \vert^q} \int _1^{\vert y\vert} r^{q + 1 - \eta q}
\, dr = \frac{1}{\vert y \vert^q}
\left(\frac{1}{\vert y\vert^{\eta q - q- 2}}-1\right). \tag3.15 $$
Hence one gets the inequality
$$F_1 \leq const \frac{1}{\vert y \vert^q} +
const \frac{1}{\vert y \vert ^{\eta q- 2}} .\tag3.16$$
Because $(p-1)q - 2 = p - 2 > 0 \, $and $(p-1)(\eta q -2) -2 = (\eta -
2)p >0$ both terms on the right hand side of 3.16 are in $L^{p-1}(\Bbb
R^2_y)$ The second term in the decomposition of $F$ is
$$F_2(y)\equiv \int _{I(y)^c}\left(f(x,y)\right)^q \,dx
= C_2 \int _{\vert x \vert \geq C_2 \vert y \vert }\frac{1}
{(1 + \vert x\vert^\eta )^q}\, dx \tag3.17$$
The integrand has no decay in y however the domain of integration
shrinks for increasing y. An explicit computation proves
$$ F_2(y) \leq const \frac {1}{\vert y \vert ^{\eta q -2}} \tag3.18$$
Hence F is again in $L^{p-1}(\Bbb R)$ , and the theorem is proved.
\qed
\enddemo
We close with the following  observations about ${Index}\, (PUP)$.
The first is a statement of stability of the relative index with
respect to deformations of the flux tube such as translating and
other local
deformations, and is a consequence of the stability of the Fredholm
index under compact perturbations.  We state one special case only:
\proclaim{Proposition (3.8)}
Let $U$ be a gauge transformation as in hypothesis (3.1)
and let $T$ be a translation, then:
$$
{Index}\, (PUP) = {Index}\,\big(P\, TUT^*\, P\big)= {Index}\,
\big(P_TUP_T\big),\quad P_T\equiv TPT^*.\tag3.19$$
\endproclaim
\demo{Sketch of Proof } $P(U-T^*UT)$ is a compact operator. This
can be seen by an adaptation of the proof of proposition 3.6 to the present
case. The stability of the index under compact perturbations gives the
first equality.  the second one follows from the invertibility of T and
the
definition of the index. \qed\enddemo
This makes the charge deficiency insensitive to
the positioning of the flux tube, (and so a global property of the
system).

There are classes of projections where the relative index is
guaranteed to vanish.  Experience with examples, such as the
quantum Hall effect, have led to the  recognition that nontrivial charge
transport  is intimately connected with breaking  time reversal  symmetry.
Indeed:
\proclaim{Theorem (3.9)} Let $U$ and $P$  satisfy hypothesis 3.1
and $P$
be time reversal  invariant, then ${Index}\, (PUP) = 0$.
\endproclaim \demo{Proof}
Since the relative index is real, 3.7 is even under conjugation. On
the other hand, time reversal invariance says that 3.7 is odd
under conjugation., so the index must vanish.  To see this, recall
that time reversal says that (in the spinless case) the integral
kernel of $P$
is real \cite{\wigner }.   It follows that the first triple product
in 3.7, $p(x,y)p(y,z)p(z,x)$, is even under conjugation. The second
triple product of 3.7,
$\big(1-\frac{u(x)}{u(y)}\big)\big(1-\frac{u(y)}{u(z)}\big)\big(1-
\frac{u(z)}{u(x)}\big)$, is odd under conjugation, since $u(x)$ is
 unimodular. It follows that the integrand in 3.7 is odd under
conjugation.
 \qed\enddemo
\proclaim{Remark (3.10)} It is easy to extend this
proof to the case of spin, and to generalized notions of time
reversal. \endproclaim
\par The next triviality result has nothing to do with time reversal,
but  rather with the geometry of $\Omega$.  It states that one can
not remove too much of $\Bbb R^2$ around the flux tube without making
the
relative index  trivial. In particular, if $\Omega$ is contained in a
wedge,
and the flux is outside $\Omega$, the index vanishes. More  precisely:
\proclaim{Theorem (3.11)} Let $U$ be a flux tube through the origin so
that $U$ and $P$ satisfy hypothesis 3.1, and let $\Omega$ be contained
in
a wedge excluding the origin,  i.e. $\Omega \subset \{z| z\in \Bbb C,
\varepsilon < arg \,  z <2\pi - \varepsilon,\ \varepsilon >0 \}$,  then,
\hbox{${Index}\,  (PUP) = 0$}.
 \endproclaim \demo{Proof} Suppose   ${Index}\, (PUP) = m$,
$m\neq 0$. Take $V\equiv U^{1/2m}$ with cut along $[0,\infty)$,
and so  entirely outside $\Omega$. Since $P$ is a projection in
$L^2(\Omega)$,  $p(x,y)=0$ if either $x$ or $y$ is in $\Omega$. Proposition
3.6 then can be  adapted to this case with  $V$ replacing $U$, using the
fact
that near the edges of the wedge  the decay in 3.1 replaces the decay
in 3.2.
It follows that $Index\,  (PVP)$ must be an integer. On the other  hand
a
little argument, using  proposition 2.5 and
Eq.\, (2.3) shows that  $m=Index\, (PUP) = 2m \, Index\, (PVP)$.
This is a contradiction. Hence  $m=0$.\qed\enddemo\vskip 0.3in \vskip
0.3in

\heading
\bf {4. Covariant Projections}
\endheading
\vskip 0.3in
In this section we  consider the relative index for projections
arising in the study of
homogeneous systems. Here we concentrate on the case of
a single Hamiltonian. In section 8 we shall consider families of
Hamiltonians with random potentials which are covariant and ergodic under
translation. The random case is of course much more interesting
from the point of view of applications to real systems. Mathematically
the case of one single covariant Hamiltonian is however the core of the
matter
as it will be seen latter.\par
The main result of this section, theorem 4.2,
gives a formula for ${Index}\, (PUP)$ which holds for projections, which,
in
addition to the assumption on the decay of their integral kernel, 3.1,
also
satisfy a condition of  covariance (or homogeneity):
Projections that are translation invariant up to a gauge
transformation.  This formula plays a key role in
relating the index to the adiabatic curvature and  Kubo's formula,
something we shall return to in the following sections. \proclaim{Definition
(4.1)} We say that a projection $P$ in $L^2(\Bbb R^n)$ is covariant
if  its integral kernel satisfies:
$$p(x,y) = \Cal U_a(x)p(x-a, y-a) \,\Cal
U^{-1}_a(y) \quad a,x,y \in { {\Bbb R}^n}.\tag4.1$$
$\Cal U_a(x)$
denotes a family of  unitary continuously differentiable
multiplication operators i.e.\ non-singular gauge transformations.
\endproclaim This notion of covariance is motivated by the covariance
for
Schr\"odinger operators with  constant magnetic  fields \cite{\zak }.
\par
It follows that the first triple product in the integrand in 3.7 is
invariant under translation of all arguments x,y,z.:
 $$\split p(x,y)p(y,z)p(z,x) &= p(x-t,y-t)p(y-t,z-t)p(z-t,x-t)
\quad t\in { {\Bbb R}^2}\\
&= p(0,y-x)p(y-x,z-x)p(z-x,0).\endsplit \tag4.2$$
This property can be used
to reduce the six dimensional integral in the computation of
${Index}\, (PUP)$ in 3.7 to a four dimensional one, provided we can
say something about two dimensional integrals with the integrand
 $\left( 1 - \frac{u(x-a)}{u(x-b)}\right)
\left( 1 - \frac{u(x-b)}{u(x-c)}\right)
\left( 1 - \frac{u(x-c)}{u(x-a)}\right)$, where $a,b$ and
$c$ are fixed points in
$R^2$.  That such integrals can be evaluated explicitly, and have
geometric significance is  a result of Connes \cite {\connes } and is
a rather amazing fact. Lemma (4.4) is in part a simplification of the
derivation and a generalization of the original observation of
Connes to the case of singular gauge transformations (Connes proof works
however also for the upper half plane).
\proclaim{ {Theorem (4.2)}}
Let $P$ be a covariant projection in $L^2(R^2)$ satisfying the decay
properties 3.1 and let  $U$ be a (singular) gauge
transformation satisfying hypothesis (3.1),  with winding
$N(U)$. Then:  $${Index}\, (PUP) = -2\pi i\,
N(U) \int\limits_{{\Bbb R} ^4}\, dx \, dy \,
 p(0,x)p(x,y)p(y,0)\, x\wedge y,\tag4.3$$
where $ x\wedge y \equiv x_1 y_2 - x_2 y_1$, $x\equiv (x_1,x_2)$
and $y\equiv (y_1,y_2)$. \endproclaim
 \proclaim{Remark (4.3)}
 The self-adjointness of $P$ gives $p(x,y)
 = \overline{p}(y,x)$,
making the Index real. If $p(x,y)$ is real the index is manifestly
$0$, as it should (by theorem 3.9). \endproclaim
The proof of the theorem needs an evaluation of an integral.
\proclaim{Lemma (4.4)}
Let $N(U)$ denote the winding number of the multiplication
operator $U$ satisfying hypothesis (3.1). Then:
$$\int\limits_{{\Bbb R} ^2}dx\,\left(
1-\frac{u(x-a)}{u(x-b)}\right) \left( 1-\frac{u(x-b)}{u(x-c)}\right)
\left( 1-\frac{u(x-c)}{u(x-a)}\right) = 2 \pi i\, N(U)
Area(a,b,c)\tag4.4$$
with $ a,b,c \in {\Bbb R}^2$ and
$Area(a,b,c) \equiv a\wedge b + b\wedge c + c\wedge a$
is twice the oriented area of the
triangle with vertices $a, b,$ and $c$.
\endproclaim
\demo{Proof}
Let
$$e(x,y) \equiv \left( \frac{u(x)}{u(y)} - \frac{u(y)}{u(x)}\right)
=-e(y,x).\tag4.5$$ Then:
$$C(a,b,c) \equiv \int\limits_{{\Bbb R} ^2} \, dx \,\big( e(x-a,
x-b) + e(x-b, x-c) + e(x-c, x-a)\big)$$
$$ = - \int\limits_{{\Bbb R} ^2}\,
dx \,\left( 1-\frac{u(x-a)}{u(x-b)}\right) \left(
1-\frac{u(x-b)}{u(x-c)}\right)
 \left( 1-\frac{u(x-c)}{u(x-a)}\right),\tag4.6$$
since the integrands of the two integrals are the same up to a
minus sign. The integral converges absolutely since each of the 3
factors can be estimated by:
$$ \left| 1-\frac{u(x-a)}{u(x-b)}\right|
\leq const |a-b| \, max\{\frac{1}{|x-a|},\frac{1}{|x-b|}\} \leq
const \frac{ |a-b|}{|x|},\tag4.7$$
for $|x| \geq const\times (|a|+|b|)$.

$C(a,b,c)$ has several manifest properties that want to make it
proportional to the oriented area of the triangle with vertices
$a,b,c$: 1. It is even or odd under cyclic or anti
cyclic permutations of $a,b,c$ respectively.
2.  It is translation invariant: $$C(a+t, b+t, c+t) =
C(a,b,c),\quad a,b,c,t \in {\Bbb R} ^2\tag4.8$$
This suggests looking at mixed second derivatives. There is a
problem however with differentiability of the integrand in the
vicinity of a,b and c and with convergence of the integral at
infinity. For that reason this bad set is cut out.
Let $B_\varepsilon(a)$ denote the ball of radius $\varepsilon$
around $a$ and let $D_\varepsilon$ be defined by:
$$D_\varepsilon \equiv B_{\frac{1}{\varepsilon}}(0) /
(B_\varepsilon(a)\cup
B_\varepsilon(b) \cup B_\varepsilon(c)).\tag4.9$$
$D_\varepsilon$ is a large disk punctured near the three points
$a,b$ and $c$ . $C(a,b,c)$ is the $\varepsilon \to 0$ limit of:
$$C_\varepsilon(a,b,c)
\equiv \int\limits_{D_\varepsilon}\, dx \,\Big(e(x-a, x-b) + e(x-b,
x-c) + e(x-c, x-a)\Big)\tag4.10$$
Since $C_\varepsilon (a,b,c)$ changes sign if two of its arguments
are interchanged, it is enough to look at the anti-symmetric second
derivatives, i.e.\ : $$\align &(\partial_{a_1}\partial_{b_2} -
\partial_{a_2}\partial_{b_1})C_\varepsilon(a,b,c)
=\int\limits_{D_\varepsilon}(\partial_{a_1}\partial_{b_2} -
\partial_{a_2}\partial_{b_1})e(x-a, x-b) \\
&=\int\limits_{D_\varepsilon}\Big( \partial_2
\overline{u}(x-b)\partial_1 u(x-a) - \partial_1
\overline{u}(x-a)\partial_2 u(x-b)\Big) - (1 \leftrightarrow 2),
\qquad \varepsilon > 0.\tag4.11\endalign$$
Using the notation of differential forms and Stokes' theorem one
gets in the limit $\varepsilon \rightarrow 0$:
$$\split (\partial_{a_1}\partial_{b_2} -
\partial_{a_2}\partial_{b_1})C_\varepsilon(a,b,c)
&=-\Big(\int\limits_{D_\varepsilon}\, d\overline{u}(x-a)\, du(x-b) -
c.c. \Big)\\
&= -\int\limits_{\partial D_\varepsilon}\Big(\overline{u}(x-b)\,
du(x-a) -c.c.\Big)\\
&\rightarrow - 4\pi \,i \, N(U).
\endsplit \tag4.12$$
The boundary $\partial D_\varepsilon$ is made of one large circle,
and three tiny circles around the puncture at $a,b$ and $c$. In the limit
$\varepsilon \to 0$ the small circles around $a,b,c$ do not contribute
to the
integral. The large circle however produces the winding number up to the
factor $2\, (2\pi i)$. \par An additional argument shows that the only
non-vanishing second derivatives of $C(a,b,c)$ are the ones just
considered (and their cyclic permutations) and that the limit $\varepsilon
\to  0$ and derivation may be interchanged.

To reconstruct $C(a,b,c)$ from its second derivatives
we integrate 4.12 twice and get:
$$C(a,b,c)= \alpha + \beta (a,b,c)- 2\pi i N(U)\,
Area(a,b,c)\tag4.13$$ where $\alpha$ is a constant
and $\beta$ a linear function. Since $C(0,0,0)=0$, we learn that
$\alpha=0$.
Since $C(a,b,c)$ and $Area(a,b,c)$ are even/odd under
permutations of $a,b,c$, so  is $\beta (a,b,c)$.  Since $\beta$ is
linear it must vanish identically. This finishes the  proof of
Lemma (4.4).
\qed\enddemo
\demo{}
Now we return to the proof of Theorem (4.2).
Using the previously introduced notation (4.6) and translational
invariance (4.2) in  (3.7) one gets:
$${Index}\, (PUP) =  \int  \, dy \, dz\,
p(0,y)p(y,z)p(z,0)\, C(0,-y,-z)\tag4.14$$
By Lemma (4.4) the proof is finished.
\qed\enddemo
\vskip 0.3in
\heading
{\bf 5. Charge Deficiency and Charge Pumps}
\endheading
\vskip 0.3in
The wave function of $n$ non-interacting fermions gives rise to a
$n-$dimensional projection in the one particle Hilbert space.
Therefore ${Index}\,  (P,Q) = {dim}\, P - {dim} \, Q$
counts the difference of the corresponding number of fermions. We
shall adopt the point of view that, with definition 2.1, ${Index}\,
(P,Q)$ also correctly   counts the difference in the number of Fermions
associated with infinite dimensional projections $P$ and $Q$.

Suppose we fix the Fermi energy in a gap in the spectrum of the
Schr\"odinger operator,  and consider the associated spectral
projection $P$. We show in Appendix A that for a wide class of
Schr\"odinger operators, the integral kernel of $P$ satisfies the decay
and regularity hypothesis in section 3.   (Presumably, these conditions
are satisfied  under weaker conditions, e.g. in the absence of an energy
gap, but provided the Fermi energy is in a region  of ``localized states").
Let $U$ be a singular gauge transformation which introduces $N(U)$ flux
quanta
into the system. $Q=UPU^*$ describes the spectral projection associated
with  the same Fermi energy, (also in a gap, by unitary invariance),
with extra  $N(U)$ units of quantum flux,  piercing $\Omega$ at
points.   Hence ${Index}\,  (P,Q)$ counts the change in the number of
electrons below the Fermi energy.

It is clear from proposition (2.5), and is manifest in Theorem  4.2, that
$Index\, (PUP)$ is linear in the number of flux quanta  carried by the
a flux tube: If the flux tube $U_1$ adds charge $q_1$  and  $U_2$ adds
charge
$q_2$, then $U_1U_2$ would  add $(q_1+q_2)$ charges.  It is therefore
natural  to define the charge deficiency  in terms of what  a flux tube
carrying one  unit of quantum flux does. And, for the sake of concreteness
we chose a  specific (rotationally symmetric) flux tube: \proclaim
{Definition (5.1)} For a  spectral projection $P$ of a  Schr\"odinger
operator
in $L^2(\Omega )$,  $\Omega \subseteq \Bbb R^2$, and  $z=x+iy$, the {\it
charge deficiency}  is  the Fredholm index $Index\, (P  \frac{z}{|z|}
P)$,
whenever the latter is  well defined.\endproclaim

In many simple cases the charge deficiency vanishes.  Proposition  3.5
tells
us that this is always the case for (reasonable Schr\"odinger operators
associated with) compact domains where the number of electrons is finite.
Nontrivial deficiency therefore requires an infinite number of Fermions.
Theorem 3.10 tells us  that even for non-compact domains with infinite
number of Fermions,  the deficiency vanishes whenever  the flux
tube is outside $\Omega$ and $\Omega$ is  contained in a wedge.  This
leaves us with infinite domains that encircle the flux tube. Finally,
even for these, theorem 3.9 tells us that the deficiency vanishes  whenever
$P$
is time reversal invariant. In particular, this is the case in the absence
of gauge fields.

It is now natural to ask whether there are examples of  Schr\"odinger
operators whose spectral projections have non-trivial deficiencies.
One way to break time reversal is with constant magnetic fields.  As we
shall see in  section 7, the simplest example of this kind, the Landau
Hamiltonian associated with the Euclidean plane,  has unit deficiency
for each Landau level.  It would be interesting to have
additional example where the deficiency is computable and non zero.
In
particular, it would be interesting to  have  examples where time  reversal
is broken in more subtle ways, for  example,
with Aharonov-Bohm fluxes.

Charge pumps are quantum mechanical devices which
transfers an integer charge in each cycle.  An interesting class of such
pumps  has been introduced by \cite{\niu }. The kind of systems discussed
in this  paper are also charge pumps.   They have a natural cycle of one
unit of quantum flux and the periodicity is exact for non-interacting
electrons. As real electrons are pumped, the pump charges. This
may modify the  effective potential in the one electron theory, and
ultimately  change the  index, destroying the periodicity.  Charging effects
are, of  course, smaller  the larger the capacitance of  $\Omega$.\par
A pump of  the kind  discussed here is stable in the sense that  deformations
in the domain $\Omega$,  the  potentials, the location of the flux tube
or the
Fermi energy would  preserve the deficiency.

To clarify the concept of charge deficency for the pair of projectors
$P$ and $Q=UPU^{-1}$ of the two Schr\"odinger operators $H$ and
$UHU^{-1}$ let us introduce a canonical interpolation between the two:
$$ H(t) = (-i\nabla - \phi (t) (\nabla \arg z) - A_0 )^2 + V, \qquad t\in
[0,1]$$
where $\phi (t) $ interpolats smoothly between zero and one.
$\nabla \arg z $  denotes a vector field on the real two plane
respectively the complex plane. $H(t)$ has, by definition, a
{\it time independent} domain of definition.
It is {\it not} unitary equivalent to  $H$ through conjugation with $U(t)
= e^{it\arg z}$ because the domain of $H$ is not invariant under $U(t)$
for t in the interior of the interval [0,1].

If we consider the time dependent dynamical system defined by the
Schr\"odinger operator $H(t)$, it is evident, that in addition to the
magnetic field $B=\nabla \times A_0$
there is an electric field $\dot \phi (t)\nabla \arg z $. It points in
the
azymuthal direction. Hence a charge experiences a Lorentz
force in radial direction and is pushed from the center of the flux
tube to infinity. This motivates the interpretation of
$P$ and $Q$ as physical states related through adiabatic dynamics of the
time dependant Hamiltonian $H(t)$ and the terminology ``charge deficiency''.

Much of the discussion above  has  analogs in the
analysis of the quantum Hall effect based on localization
of wave functions \cite {\laughlin ,\ka ,\pg ,\thouless }.
\vskip 0.3in
\heading
{\bf 6. Adiabatic Curvature and Hall Conductance}
\endheading
\vskip 0.3in
In this section we discuss the Hall charge transport, which
is a priori distinct from charge deficiency discussed in
previous sections.   This notion is related to adiabatic curvature,
Chern numbers, and to Kubo's formula.  We describe this in some
detail.  The main result, theorem 6.6,  says that under appropriate
conditions the Hall charge transport and charge deficiency are
related.

As in our discussion of charge deficiency, we consider a
cycle of  Schr\"odinger operators associated with a
gauge transformation. However, the  gauge transformation
is not associated with a flux tube that pierces the system.
Rather, it is associated with a (finite) voltage drop across the
system whose time integral is a unit of quantum flux.  This voltage drop
is
associated with a class of functions, which we call {\it switches} and
which, roughly, look like the graphs of $\frac{1}{2}\tanh (x)$. More
precisely:
\proclaim {Definition (6.1)}
$ \Lambda (x)$, $x \in \Bbb R$, a function of one variable, is called
a switch
if it is a continuously differentiable, real valued, monotone, non-
decreasing function such that the limits at $+\infty$ and $-\infty$ exist
and
$$\Lambda  (\infty ) - \Lambda (-\infty ) = 1.     \tag6.1$$
\endproclaim The
setting relevant to this section is described in the following:
\proclaim{Hypothesis (6.2)} Consider the family of, unitarily
equivalent,
magnetic Schr\"odinger operators in $L^2(\Bbb R^2)$,
$$\eqalign{H(A,V)\equiv
(-id -A)^2 + V &= e^{i\, (\Phi_1 \Lambda _1+\Phi_2 \Lambda
_2)}\Big((-id
-A_0)^2 + V\Big)e^{-\, i \,(\Phi_1 \Lambda _1 +\Phi_2 \Lambda
_2)},\cr A
&\equiv A_0 + \Phi_1 \, d\Lambda _1 + \Phi_2 \,
d\Lambda_2,}\tag6.2$$
where:\par
a) $A_0$ and $V$, the vector and scalar potentials, satisfy the
(mild) regularity conditions in Appendix A;  $\Phi \equiv (\Phi _1,\Phi
_2)\in \Bbb R^2$ and  $ \Lambda _1,\Lambda_2$ are both switches.
\par
b) $$P(\Phi) = e^{i\, (\Phi_1 \Lambda _1
+\Phi_2 \Lambda _2)} P(0)e^{-\, i \,(\Phi_1 \Lambda _1 +\Phi_2
\Lambda _2)}.\tag6.3)$$
 is  a family of spectral projections for $H(A,V)$ associated with a
Fermi
energy in a gap in the spectrum.\endproclaim
\proclaim {Remarks }
1. In Appendix A we
show that b)  of hypothesis 6.2 implies that the integral kernel of
$P$
satisfies the regularity and decay properties in hypothesis
3.1.\newline
2. In the case where $\Phi$ is time dependent, $\dot\Phi_1$ is the
voltage drop along the x-axis and   $\dot\Phi_2$ is the voltage
drop along the y-axis.\newline 3.  The monotonicity condition on
the switch functions  implies integrability of the derivative of switches
in
the absolute sense and  enters in the proof of proposition 6.9.
\endproclaim
We recall:
\proclaim{Definition (6.3)}
 The adiabatic curvature
associated to $P$ is: $$\omega _{12} \equiv i \, P\left[ \partial
_{\Phi_1}P,\partial _{\Phi_2}P\right]P.\tag6.4$$
\endproclaim
A direct calculation gives: $$\omega _{12}
= - \, i \, [P\Lambda _1 P, P \Lambda _2 P] = \, i \, P[\Lambda _1
P_{\bot} \Lambda _2 - \Lambda _2P_{\bot} \Lambda _1]P
= i\Big( [P,\Lambda_1]P_\bot  [P,\Lambda_2] - (1
\leftrightarrow 2)\Big).\tag6.5$$ Furthermore, since
$\Lambda _1$ and
$\Lambda _2$ are multiplication operators: $$\omega _{12} (\Phi)
= e^{i\,
(\Phi_1 \Lambda _1 +\Phi_2 \Lambda _2)} \omega_{12}(0)e^{-\, i
\,(\Phi_1 \Lambda _1 +\Phi_2 \Lambda _2)}.\tag6.6$$
It would be nice, if hypothesis 6.2 were to imply that  the adiabatic
curvature  is trace class.  Since we do not know if this is the case,
we shall study  traces by taking limits.  To this end we introduce:
\proclaim{Notation (6.4)} Let $\Omega \subset{\Bbb R} ^2 $
denote the square box $[-L,L]\times [-L,L]$, and let $\chi
_\Omega$ be the characteristic function of the box.  $|\Omega |$
denotes the area of the box.
\endproclaim
The unitary equivalence of the family in 6.2, makes the adiabatic
curvature $\Phi$ independent in the following sense:
\proclaim{Proposition (6.5)}
Let $P$ be a spectral projection associated with a gap, then $\chi
_\Omega\omega _{12} \chi _\Omega$ is trace class and its trace is
independent of $\Phi$. \endproclaim
\demo{Proof}
 Since $\Lambda _1 P_{\bot} \Lambda
_2 - \Lambda _2 P_{\bot} \Lambda _1$ is
bounded it is enough to prove that $\chi _\Omega P$ is Hilbert-
Schmidt
(recall that all Schatten classes are ideals). By the theorem in
Appendix A  the integral kernel of $P$
satisfies the decay properties  (3.1).  Consequently,
$$\int \, dx \,
dy \, |\chi _\Omega (x) p(x,y)|^2 < \infty \tag6.7$$ The
$\Phi$-independence is obvious from (6.6).
\hbox{\hfil$\square$}
\enddemo
For our  purpose, the most convenient way of
introducing charge transport in the Hall effect is to  {\it define} it
by:
\proclaim{Definition (6.6)}
 The Hall charge transport, $Q$, is
$$ Q
\equiv -2 \pi \,\lim_{L\to \infty} {{Tr} \,\chi _\Omega \,\omega
_{12}
\chi _L}.\tag6.8$$
\endproclaim
\proclaim{Remarks (6.7)}
a) Theorem 6.8 below guarantees the existence of the limit,
under the conditions in Hypothesis 6.2. \newline b)  In our
units, the Hall {\it conductance} is $Q/2\pi$.\newline c) Our sign
convention is such that the Hall conductance of a full Landau
level is $1/2\pi$.\endproclaim \par The physical interpretation of
charge transport introduced here is the following. It  is the charge
that crosses the $x_1$ axis, in the positive direction, as the Hamiltonians
in 6.2
undergo a cycle  corresponding to adiabatically increasing $\Phi_1$
from $0$ to $2\pi$.   (Alternatively, it is minus the charge that crosses
the $x_2$ axis as the Hamiltonians in 6.2 undergo a cycle
corresponding to adiabatically increasing $\Phi_2$  from $0$ to
$2\pi$). This is the transport in the Hall effect. For more on this the
the reader may want to consult \cite{\asy ,\ka,\ks  ,\nt ,\ntw }.

The following theorem is the central result of this section. It
says that the Hall conductance can sometimes be interpreted as
an index. The strategy is to show that definition 6.6 can be put
 into the form of Theorem 4.2.
 \proclaim{\bf Theorem (6.8)}
Suppose $P$ is a covariant projector, $P$ and $P, \Lambda
_{1,2}$ satisfy the hypothesis 6.2. Then the Hall charge transport
$Q$
equals the charge deficiency:
$$Q = -2 \pi \, i \,\int dy \, dz \,
p(0,y)p_{\bot} (y,z) p(z,0)\, y \wedge z =  -{Index}\left(P
\frac{z}{|z|}P\right) .\tag6.9$$ \endproclaim

The proof of the theorem, like  that of theorem 4.2 depends on
an explicit evaluation of (another) area integral  and this
one too is related to areas of triangles. We start with this
preparatory proposition: \proclaim{\bf Proposition (6.9)}
For  $\Lambda$  a switch
$$
 \int_{\Bbb R}dx (\Lambda(x+a)-\Lambda(x))=a,\qquad a\in \Bbb
R.\tag6.10 $$
If both  $\Lambda_1$ and $\Lambda_2$ are switches, then
$$  \int_{\Bbb R^2} dx_1 dx_2\,
\Big(\big(\Lambda_1(x_1+a_1)
-\Lambda_1(x_1)\big)\big(\Lambda_2(x_2+b_2) -
\Lambda_2(x_2+a_2))- \big(1 \leftrightarrow 2\big)\Big) =
a\wedge
b ,\tag6.11$$ where $ a\wedge
b\equiv  a_1b_2 -a_2b_1$. Both integrals converge absolutely.
\endproclaim
\demo{Proof}
a) Look at
$$\align
\int\limits_{-\infty}^\infty dx\,\big(\Lambda(x+a)-
\Lambda(x)\big)
&= \int\limits_{-\infty}^\infty dx\int_x^{x+a}dt\,\Lambda'(t)
= \int\limits_{-\infty}^\infty dx\int_0^{a}dt\,\Lambda'(t+x)\\
&= \int_0^{a}dt\int\limits_{-\infty}^\infty
dx\,\Lambda'(t+x)=\int_0^a dt =
a. \tag6.12\endalign$$ Monotonicity of the switch implies absolute
convergence. \newline
b)  From, 6.10
$$  \int_{\Bbb R^2} dx_1 dx_2\,
\big(\Lambda_1(x_1+a_1)
-\Lambda_1(x_1)\big)\big(\Lambda_2(x_2+b_2) -
\Lambda_2(x_2+a_2)\big)= a_1(b_2-a_2).\tag6.13$$
And similarly with $1 \leftrightarrow 2$.  Subtracting the two gives
6.11.
\qed
\enddemo

\demo{Proof of Theorem 6.8}
To compute the transport according to definition 6.4 we look first
 at the integral kernel of the adiabatic curvature (the last identity
in 6.5) restricted to the diagonal
$$\omega_{12}(x,x) = i\int_{\Bbb R^4}dy\,dz\,p(x,y)\,p_\bot
(y,z)\,p(z,x)
\Big(\big(\Lambda_1(y_1) -
\Lambda_1(x_1)\big)\big(\Lambda_2(z_2) -
\Lambda_2(y_2)\big)-
\big(1 \leftrightarrow 2\big)\Big).\tag6.14$$
Due to translational invariance the integrand in 6.14 can be
replaced by
$$i \, p(0,y)\,p_\bot (y,z)\,p(z,0)
\Big(\big(\Lambda_1(y_1+x_1)
-\Lambda_1(x_1)\big)\big(\Lambda_2(z_2+x_2) -
\Lambda_2(y_2+x_2)\big)- \big(1 \leftrightarrow
2\big)\Big).\tag6.15$$
To compute the charge transport we have to integrate the above
expression over the domain $\Omega$ and after that let $L\to \infty$.
\,
Since  all integrations converge absolutely even for $\Omega = \Bbb R^2$
we are permitted to exchange the order of integration and the limit $L
\to \infty$. Hence we integrate first over x, then we let
$L \to \infty$ and then we integrate over $y$ and $z$. The x
integration can be done by b) of proposition 6.9. Putting this into
the definition of  the Hall
charge transport $$ \align Q &= -2\pi i \int_{\Bbb
R^4}dy\,dz\,p(0,y)\,p_\bot (y,z)\,p(z,0)\,
 y\wedge z \\ &= 2\pi i \int_{\Bbb
R^4}dy\,dz\,p(0,y)\,p(y,z)\,p(z,0)\,
 y\wedge z.\tag6.16\endalign$$
This proves the first part of the theorem. The second part is a
consequence of theorem 4.2.
\qed\enddemo

To relate this expression to Kubo's formula is
rather simple. We start from 6.9, multiplying the integral
formula  by $1= \frac{1}{|\Omega|}\int\limits_{\Omega} dx$, and
use the covariance of the projectors (4.3) to get :
 $$ Q =-   \frac{2 \pi \, i}{|\Omega|} \int_{\Omega} dx \,
\int_{\Bbb R^4}
dy dz  p(x,y)p_{\bot} (y,z) p(z,x) {(y-x) \wedge (z-x)}.\tag6.17 $$
The terms arising from terms linear and quadratic in $x$ again
vanish. Hence, the conductance,
 $$\eqalign{\frac{Q}{2\pi}& =- { \frac{i }{|\Omega
|}}\int\limits_{\Omega}
dx \,\int_{\Bbb R^4} dy\, dz\,    p(x,y)p_{\bot} (y,z) p(z,x) {y
\wedge z}\cr &=
 - { \frac{ i }{|\Omega|} }{Tr}\,\big(
\chi _\Omega (Px_1 P_{\bot} x_2 P - P x_2 P_{\bot}  x_1 P)\big),}
\tag6.18$$
which is Kubo's formula.
\vskip 0.3in
\heading
{\bf 7. Landau Hamiltonians}
\endheading
\vskip 0.3in
It is instructive to consider an example where the theory of the
previous section applies and, moreover, is non trivial in the sense that
it
gives non-zero deficiency. Such an example is provided by  Landau
Hamiltonians and the spectral projections on Landau levels. The Landau
Hamiltonian in $L^2 ({\Bbb R} ^2)$ is: $$H(A) \equiv \frac {1}{2}(-\,
i\,d - A)^2\tag7.1$$
where $dA = B\, dx\wedge dy$. $B>0$ is a constant magnetic field.
$Spectrum \,\big(H(A)\big) = \{ \frac {1}{2} B (2n+1)\, | n\in
\Bbb N\}$, and  each point in the spectrum, a Landau level, is
infinitely degenerate. We shall denote the spectral projection on the
n-th
Landau level by $P_n$. Clearly, $dim\, P_n = \infty $.  We show below
that
projections on Landau levels satisfy  hypothesis 3.1, and that the charge
deficiency of each Landau level is unity.
\proclaim{Proposition (7.1)} Let $H(A)$ be the Landau Hamiltonian
with $B>0$, $A$ differentiable and  $P_n$ the projection on the $n$-th
Landau
level. Then $p_n (x,y)$ is covariant, jointly continuous in $x$ and
$y$, and decays like a gaussian in the  variable $|x-y|$. In particular,
hypothesis (3.1)
holds. \endproclaim \demo{Proof} a) Let $T_a$ denote the
translation by
$a\in \Bbb R^2$. Since $B$ is constant and $\Bbb R^2$ is simply
connected, $A(x-a) -A(x) = d \Lambda_a(x) = i \,\Cal U_a^* d \Cal U_a$
with
$\Cal U_a(x) \equiv \exp -i\,\Lambda_a(x)$.  It follows that
$$T_a \,H(A) T_{-a} = (-\, i\,d - A(x-a))^2= (-\, i\,d - A(x) -
d\Lambda_a(x))^2= \Cal U_a^*\, H(A)\, \Cal U_a.\tag7.2$$
Hence $H(B)$ commutes
with magnetic translations $\Cal Z_a  \equiv \Cal U_aT_a$  \cite
{\zak }. The spectral projections are covariant in the sense of section
4 and
$$p(x,y) = \Cal U_x^{-1} (x)\,  p(0,y-x) \,\Cal U_x (y).\tag7.3$$
b) With $A$ and $A'$ related by a (continuous) gauge
transformation
$\Lambda$, $A' = A+ d\Lambda $, the corresponding integral
kernels are
related by $p_{A'} (x,y) = e^{i\,\Lambda(x)} p_A(x,y)e^{-
i\,\Lambda(y)}$, and
so $p_{A'}(x,y)$ is continuous in $x$ and $y$ if $p_A(x,y)$ is.  It is
therefore enough to check the regularity and decay for a specific choice
of
$A$. By scaling the coordinates, we may take $B=2$.  We shall now show
that for  $A_0 \equiv \frac{1}{2} (-y\, dx + x\, dy)$, $p_0 (0,z)=
Polynomial (z)\,\exp -|z|^2/2$.  Which proves the regularity and decay.
The
corresponding  the Landau Hamiltonian is:
$$H(A_0) = 2 D^* D + 1,\quad D\equiv (\partial_{{}\overline{z}} + \frac{z}{2}
),
\quad z = x + iy.\tag7.4$$
The lowest Landau level is spanned by:
 $$<z|n,0>\, = (\pi n!)^{-1/2} z^n e^{-|z|^2 / 2},\quad n = 0,1,\dots\tag7.5$$
and the $m$-th
Landau level by $$<z|n,m>\, = (\pi n ! (m+1)!)^{-1/2} (D^* )^m (z^n
e^{-|z|^2/2}).\tag7.6$$ Since $<0|n,m> = 0$ for $m \neq n$ we have:
$$p_m(0,z) =\sum_n <0|n,m><n,m|z>\, =\, <0|m,m><m,m|z>\tag7.7$$ which
is
smooth andwith gaussian decay. \qed\enddemo It follows that the results
of
the previous sections apply. In particular,  the deficiency is a finite
integer and the Hall conductance for the n-th Landau level is   $-
\frac{1}{2\pi}\, Index\,\left(P_n \frac{z}{|z|} P_n\right)$.  It remains
to
compute  the
index.  This computation depends on the following simple lemma:
\proclaim {Lemma (7.2)} Let $M$ be a semi-infinite Fredholm
matrix so
that its non-zero entries lies on the i-th sub-principle diagonal, i.e.:
$$\big( M\big)_{mn}= c_m \,\delta_{m+i,n},\quad n,m\in \Bbb
N,\quad i\in \Bbb Z,\tag7.8$$ then, $Index\, M = i$.\endproclaim
\demo {Proof} Suppose first that all the $c_m \neq 0$. The kernel
of $M$
is spanned by the  projection on the first i dimensions.
The kernel of $M^*$ is empty.  Consequently $Index\, M =i$.  Now
to the general case: Since $M$ is Fredholm there is at most a finite
number of $c_m=0$. Deforming a finite number of $c_m$ to zero, does not
change the
index by the stability under compact perturbations, and so
$Index\, M =i$.
\qed\enddemo
That the Hall conductance of each full Landau level is $1/2\pi$ is
known from  $1001$ different calculations and arguments.  The following
computation, via an index, gives the 1002 way of seeing that:
\proclaim{Proposition (7.3)} For the m-th  Landau levels: $${Index}
\,(P_m \,
\frac{z}{|z|} P_m) = -1.\tag7.9$$ In particular, the charge transport
and charge deficiency of each Landau level is unity. \endproclaim
\demo{Proof}
From (7.6)  one sees that the state $<z|n,m>$ has angular momentum
proportional to $n-m$. Consequently, the matrix elements of $\left(
P_m
\frac{z}{|z|} P_m \right) _{n, n'}$ are: $$\left( P_m \frac{z}{|z|} P_m
\right)
_{n, n'} = \delta _{n,n'+1} c(m,n). \tag7.10 $$  The result now
follows from lemma 7.2.\qed\enddemo  \par As we have discussed
in previous sections, the charge deficiency may be thought of as the
change of number of electrons in a cycle where a flux tube carrying one
unit
of quantum flux is introduced into the system. In the present situation
one
can  follow this cycle by the spectral analysis of the Landau Hamiltonian
with a  flux  tube carrying any real flux.  One finds that as the flux
increases
by  one unit,
$n$ states from the n-th Landau level descend to the $n-1$ Landau
level, and one state is lost to infinity \cite{\ap ,\laughlin }. \vskip
0.3in

\heading {\bf 8. The Ergodic Case} \endheading \vskip 0.3in In this
last section we extend the results of sections 4  and 6 about covariant
families of projectors to the case of an ergodic family of Schr\"odinger
operators, $H(A, V_\omega)$: $\omega$ is a point in probability space
$\tilde\Omega$,
the action of translations on $\tilde\Omega$ is ergodic and:
$$V_\omega
(x+a) = V_{T_a \omega} (x) \tag8.1$$ We shall denote integrals with
respect to the  probability measure  by $< \cdot >$. This family of
Schr\"odinger operators is one of the canonical  models
for the integer Hall effect. \proclaim{Proposition (8.1)}Let
$P_\omega$ be a spectral projection for $H(A, V_\omega)$ satisfying
hypothesis 3.1,\,\,$\omega \in \Omega$. Then ${Index}\, (P_\omega
UP_\omega)$ is measurable with values in $\Bbb Z$. In fact
${Index}\, (P_\omega  UP_\omega)$
is integer and constant almost everywhere .
\endproclaim \demo{Proof}
We prove first that ${Index}\, (P_\omega UP_\omega)$ is
measurable. Due to proposition 2.2 and 2.4 the index can be expressed
in terms of a trace
$${Index}\, (P_\omega UP_\omega) =  {Tr}\,  (P_\omega-
Q_\omega)^{2n
+1},
\qquad Q_\omega \equiv U_\omega P_\omega U_\omega^{-1}.\tag
8.2$$
Hence it is enough to prove measurability of the operator as an
operator valued function of $\omega$, i.e.\ measurability of the
scalar product
$(f,(P_\omega- Q_\omega)^{2n +1}f)$, $ f \in L^2(\Bbb R^2)$. But the
resolvent and therefore the projector $Q_{\omega}$, which by
assumption can be expressed in terms of an integral over the
resolvent, is  measurable. This proves the assertion.

Secondly. the  function $I(\omega) \equiv {Index}\,  (P_\omega U
P_\omega)$ takes integer values. Hence $\tilde\Omega = I^{-1}
({\Bbb Z})$. Furthermore for every $k \in ({\Bbb Z})$,\,
$I^{-1} (k)$
is an invariant set in $\tilde\Omega$ under the action of
translations. This is seen as
follows: Let $\Cal Z_a$ denote again the magnetic translation. Since
$$(\Cal Z_a\, V_\omega \,\Cal Z^{-1}_a) (x) = V_\omega (x-a)
= V_{T_a \omega}(x),\tag8.3$$
we have:
$$\Cal Z_a\, P_\omega \,\Cal Z^{-1}_a = P_{T_a \omega} \tag8.4$$
Since the index is shift invariant (Proposition (3.8)) we have:
$${Index}\, (P_\omega \, U \, P_\omega)
 = {Index} \, (P_{T_a \omega}\, U \, P_{T_a  \omega}).\tag8.5$$
So the index is constant on the orbits of translations. Due to
ergodicity, the measure of $I^{-1} (k)$,
is zero or one for all $k \in {\Bbb Z}$, . Since
$$\mu (\tilde\Omega) = 1 = \sum_{k \in {\Bbb Z}} \mu\left( I^{-1}
(k)\right)\tag8.6$$
it follows that there is just one $k_0 \in {\Bbb Z}$
 for which $\mu\left(
 I^{-1} (k_0)
\right) = 1$.
\qed\enddemo
In the ergodic situation the analog of (4.1) is:
$$P_\omega (x,y)
= \Cal U_a (x) P_{T_a \omega} (x-a, y-a) \Cal U_a^{-1} (y)
 \tag8.6$$
This means that the analog of (4.2) is: The
triple product that enters the basic formula, 3.7,
$$<P_\omega (x_1, x_2)
P_\omega (x_2, x_3) P_\omega (x_3, x_1)>,\tag8.8$$
is translation invariant
i.e. it does not change under the substitution $x_i \rightarrow x_i +
a$,
\,$\omega \rightarrow T_a \omega$, $a \in {\Bbb R}^2$.

We see that we get an analog of Theorem (4.1) at the price of
averaging over probability space. Namely,
 \proclaim{\bf Theorem (8.2)}
Let $H(A,V_\omega)$ be a family of ergodic Schr\"odinger
operators and $U$  a unitary operator with unit winding number
satisfying hypothesis 3.1 for all $\omega \in \Omega$, in particular
$p_{\omega}(x,y)$ satisfies inequality 3.1. Then
the average Hall charge transport $<Q>$ satisfies, a.e.:
$$<Q> = - {Index}\, (P_\omega U P_\omega)\tag8.9$$
\endproclaim
\demo{Proof}
The proof  of this statement is an adaptation of the one given in
section 4, theorem 4.2; integrating the basic equality 3.7 for the
index over probability  space brings us into the
situation we had encountered in the proof of theorem 4.2 since the
average of the triple product (8.8) is invariant under translations.
\qed\enddemo
\vskip 0.3in
\heading
{\bf Appendix A}
\endheading
\vskip 0.3in
The purpose of this appendix is to show that hypothesis 3.1 on the
regularity and decay of the integral kernel of spectral projections is
guaranteed whenever the Fermi energy is placed in a gap. Although
we have not attempted to give optimal conditions on the vector
potentials, the conditions are mild enough to cover the physically
interesting models.
\proclaim {Theorem (A.1)} Let $H(A,V)$ be a one particle
Schr\"odinger operator in  $n=2,3$ dimensions with differentiable vector
potential $A$  and scalar  potential
$V$ which is in the Kato class $K_{n=2,3}$ (which includes
Coulombic singularities).\newline a) The integral kernel for spectral
projections for $H(A,V)$, $p(x,y)$   is jointly continuous in $x$ and
$y$.\newline b) Suppose, in addition, that $H(A,V)$ has a gap in the
spectrum.  Then   the spectral projection below the gap has integral kernel
which  decays exponentially with $|x-y|$. \endproclaim \proclaim{ Remark
}
The  two parts of the theorem have rather different proofs. The $K_n$
condition is natural for (a). Part (b) only requires form boundedness
of $V$
which is slightly weaker than the $K_n$ condition.
\endproclaim\demo{Proof} {(a)} $\exp (-tH)(x,y)$ has a jointly
continuous integral kernel by the path integral (Ito) way of writing the
kernel  -- see, e.g.\ \cite{\semigroup }.  Because $H$ is bounded below
and
has a  gap, $P = g(H)$ where $g$  is a smooth function of compact support.
Since  $f(y) \equiv  \exp (2y)  g(y)$ can be approximated by polynomials
$\exp  (-y)$ uniformly, we can  write
       $$ g(H) = lim\,  g_j (H),\quad
        g_j(H) \equiv \exp(-H) f_j (H) \exp(-H),\eqno(A.1)$$
where the operators $f_j$ converge to $f$ in norm as
$L^2\rightarrow L^2$ operators and each $f_j(H)$ is a polynomial in
$\exp (-H)$.  On  general principles (see, e.g.\,  \cite{\semigroup }),
$\exp
(-H)$ is a bounded  operator from $L^1$ to $L^2$ and from $L^2$ to
$L^\infty$.  Thus
the limit in  (A.1) gives a bounded operator from $L^1$ to
$L^\infty$ and so in
infinity norm  for the integral kernel (see e.g. \cite{\semigroup }).
Since
$g_j$ has a  continuous integral kernel the result follows.
\newline {(b) } Let $B_{\vec a} \equiv e^{i\,\vec a \cdot \vec x}$, $a
\in {\Bbb C}$, be a complex boost. Then: $$B_a\, H(A,V)\, B_{-a} = H(A,V)
+
\vec a \cdot \vec a  + \vec a \cdot (-i\,\vec \nabla -\vec A).\tag A.2$$
This gives an analytic family of
type $B$ in the sense of Kato \cite{\kato } if the form domain is
independent of $\vec a$. In particular, this is the case if $V$ is form
bounded relative to  the kinetic energy. By the diamagnetic inequality
it is
enough to check  that $V$ is bounded relative to the Laplacian.  $K_n$
implies form boundedness (see \cite{\semigroup } ). In particular, if
$P$ is
a spectral projection associated with a gap, then  the
gap is stable and: $$p_a (x,y) = e^{i\, a \cdot x} p (x,y) e^{-i\, a \cdot
y}\tag A.3$$ is real analytic in $\vec a$ uniformly in $x$ and $y$. In
particular, (A.2) says that $p(x,y)$ is exponentially decaying in $|x-y|$.
This
is  a version of the Combes--Thomas argument \cite{\ct }. \qed\enddemo
\proclaim{Remarks } 1. For potentials $V$ which are perturbations of Landau
Hamiltonian, an adaptation of the above method gives decay which
is faster than  any exponential.\newline 2. It is easy to construct families
of
Schr\"odinger operators,  with ergodic $A$ and $V$ so that
$H(A,V)$ has gaps in the spectrum. \newline 3. A central open question
is
wether  the integral kernel of spectral projections for ergodic
Schr\"odinger  operators in two dimensions automatically satisfy the decay
assumption of hypothesis 3.1 for most Fermi energies.\endproclaim \vskip
0.3in  \vskip
0.3in \heading {\bf  References}
\endheading
\vskip 0.3in
\item{\ap .} J.~E.~ Avron and A.~Pnueli, ``Landau Hamiltonians
on Symmetric Spaces", in {\it Ideas and Methods in Mathematical analysis,
stochastics, and
applications Vol II}, S.~Albeverio, J.~E.~Fenstad, H.~Holden and
T.~Lindstr\o m, Editors, Cambridge University Press, (1992).
\item{\arz .} J~.E.~Avron, A.~Raveh and B.~Zur, ``Adiabatic quantum transport
in  multiply connected systems'', Reviews of Mod. Phys. {\bf 60},  873-916,
(1988).
\item{\asy .}  J~.E~. Avron, R~. Seiler and L~.G~.
Yaffe,  ``Adiabatic theorems and applications to the quantum Hall effect",
 Comm.~Math.~Phys.~{\bf 110},  33-49, (1987); J~.E~. Avron and R~. Seiler,
"Quantization of the Hall Conductance for General Multiparticle Schr\"odinger
Hamiltonians", Phys.Rev. Lett. 54, 259-262, 1985.
\item{\ass .} J~.E~. Avron, R~. Seiler and B.~Simon, ``The  index of a
pair of projections'', in preparation.
\item{\bellissard .} J.~Bellissard,``Ordinary quantum Hall effect
and non-commutative cohomology'', in
{\it Localization in disordered systems},
W.~Weller \& P.~Zieche~ Eds., Teubner, Leipzig, (1988).
\item{\birman .} M.~Sh.~Birman, ``A proof of the Fredholm trace formula
as an application of a simple embedding for kernels of
integral operators of trace class in $L^2 (\Bbb R^m )$"
Preprint, Department of Mathematics, Linkping University,
S-581 83 Linkping, Sweden.
\item{\bw .} B.~Block and X.~G.~Wen, ``Effective theoreis of the
Fractional quantum Hall effect at generic filling fractions'', \prb {\bf
42}, 8133--8144, (1990) and ``Effective theoreis of the
Fractional quantum Hall effect: Hierarchy construction'', {\it ibid}.
8145-8156 and ``Structure of the microscopic theory of the hierarchical
fractional quantum Hall effect'', {\it ibid}. {bf 43}, 8337--8349, (1991).
\item{\bmm .} M.~Bregola, G.~Marmo and G.~Morandi, Eds. {\it  Anomalies,
Phases, Defects,\dots}, Bibliopolis, (1990). \item{\carey .} A.~L.~Carrey,
``Some homogeneous spaces and representations of the Hilbert Lied group'',
Rev.\ Rom.\ Math.\ Pure.\ App.\ {\bf 30}, 505-520, (1985).
 \item{\ct .} J.~M.~Combes and L.~Thomas, 
``Asymptotic Behavior of Eigenfunctions
for Multiparticle Schr\"odinger Operators'' \cmp \
{\bf 34}, 251--270, (1973). \item{\connes .} Alain Connes, ``Noncommutative
differential  geometry'', Pub.
Math. IHES, {\bf 62} 257--360 (1986); and {\it  Geometrie Non
Commutative}, InterEdition, Paris, (1990).
\item{\cuntz .} J.~Cuntz ``Representations of quantized
differential forms in non-commutative geometry", in {\it Mathematical
Physics X}
K.~Schm\"udgen
Ed., Springer, (1992).
\item{\cfks .} H.~L.~Cycon, R.~G.~Froese, W.~ Kirsch,  and
B.~Simon {\it  Schr\"odinger Operators}, Springer,   (1987).
\item{\dn .} B.~A.~Dubrovin and S.~P.~Novikov, Sov. Phys. JETP  {\bf
52}, 511,
(1980).
\item{\efros .} E.~G.~Efros,``Why the circel is connected", Math.\
Intelligencer, {\bf 11}, 27--35, (1989).
\item{\fedosov .} B.~V.~Fedosov, ``Direct proof of th formula for the
index of
an elliptic system in Euclidean space'', Funct.\ Anal.\ App.\ {\bf 4},
339--341, (1970).
\item{\fradkin .} E.~Fradkin, {\it Field theories of condensed matter
systems}, Addison-Wesley, (1991).
\item{\f .}  J.~Fr\"ohlich and T.~Kerler, "Universality in Quantum Hall
Systems",
Nucl.~Phys.~{\bf B354}, 369,
(1991); J.~Fr\"ohlich and U.~Studer, ``Gauge Invariance in Non-Relativistic
Many Body Theory'', in {\it Mathematical Physics X},
K.~Schm\"udgen Ed., Springer, (1992).
\item{\hormander .} L.~H\"ormander, ``The Weyl calcuclus of
Pseudo-Differential operators'', Comm.\ Pure and\ App.\ Math.\
{\bf 18},
501-517,  (1965).
\item{\ka .} H.~Kanamura and H.~Aoki, {\it The physics of
interacting electrons in disordered systems}, Clarendon Press,
(1989).
\item{\kato .} T.~Kato,  {\it Perturbation Theory for Linear
Operators}, Springer, (1966).
\item{\kg .} A.~A.~Kirillov and A.~D.~Gvishiani, {\it
Theorems and problems in Functional Analysis}, Springer, (1982).
\item{\ks .} M.~Klein and R.~Seiler, ``Power law corrections to the Kubo
formula vanish in quantum Hall systems'', \cmp {\bf 128}, 141, (1990)
\item{\kohmoto .} M. Kohmoto, "Topological invariants and the quantization
of the Hall conductance'', Ann.\ Phys.\ {\bf 160}, 343-- 354, (1985).
\item{\kunz .} H.~Kunz, ``The quantum Hall effect for electrons in
a random potential",  \cmp \ {\bf 112}, 121, (1987).
\item{\laughlin .} R.~G.~Laughlin,
 ``Elementary theory: The incompressible quantum fluid'',  in:
{\it The Quantum Hall Effect},
 R.~E.~Prange and S.~M.~Girvin, Eds., Springer, (1987).
\item{\matsui .} T.~Matsui, ``The Index of scattering operators of
Dirac equations",\cmp \ {\bf 110}, 553--571, (1987).
\item{\nb .} S.~Nakamura and J.~Bellissard, ``Low bands do not contribute
to quantum Hall effect'', \cmp {\bf 131}, 283-305, (1990).
\item{\niu .}  Q. Niu,  ``Towards a quantum pump of electron
charge'', Phys.\
Rev.\ Lett.\   {\bf 64}, 1812,  (1990); ``Towards an electron load
lock'', in {\it
Nanostructures and Mesoscopic Systems}, W.~P.~Kirk and
M.~A.~Reed eds.,
A.P.\ (1991).
\item{\nt .} Q.~Niu and  D.~J.~Thouless, ``Quantum Hall effect with
realistic
boundary conditions,'' \prb \  {\bf 35},
2188, (1986).
\item{\ntw .} Q.~Niu, D.~J.~Thouless and Y.~S.~Wu,
``Quantum
Hall conductance as a topological invariant",\prb \ {\bf 31}, 3372--
3379,
(1985).
\item{\pg .} R.~E.~Prange and  S~.M.~Girvin, {\it The Quantum Hall
Effect} ,
Springer, (1987)
\item{\russo .} S.~Russo,  ``The norm of the $L^p$ Fourier transform on
unimodular groups'', Trans. AMS, {\bf192}, 293--305, (1974) and ``On the
Hausdorff-Young theorem for integral operators'', Pac.\ J.\ Math.\ {\bf
28},
1121--1131, (1976).
\item{\seiler .} R.~Seiler, ``On the quantum Hall
effect", in {\it Recent developments in Quantum Mechanics}, A.\ Boutet
de Monvel et.\ al.\ Eds., Kluwer, Netherland, (1991).
 \item{\sw .} A.~Shapere, F.~Wilczek,{\it
Geometric Phases in Physics}, World Scientific, Singapore, (1989)
\item{\trace .} B.~Simon, {\it Trace Ideals and their  Applications},
Cambridge Uni. Press, (1979).
\item{\semigroup .} B.~Simon, ``Schr\"odinger
Semigroups'', Bull.~AMS  {\bf 7}, 447--526, (1982).
\item{\stone .} M.~Stone, Ed. {\it Quantum Hall effect}, World
Scientific, (1992).
\item{\streda .} P.~\v Streda, "Theory of quantized Hall conductivity
in two dimensions", J. Phys. C {\bf 15}, L717, (1982).
\item{\tknn .} D.~J.~Thouless, M.~Kohmoto,~P.~Nightingale and M.
den Nijs, ``Quantum Hall conductance in a  two dimensional periodic
potential",
Phys.~Rev.~Lett. {\bf 49}, 40, (1982).
\item{\thouless .} D.~J.~Thouless,``Quantisation of particle
transport'',
\prb \ {\bf 27}, 6083, (1983).
\item{\wen .} X.~Wen, "Vacuum degeneracy of chiral spin states
in compactified space",\prb {\bf40}, 7387-7390, (1989); "Gapless
boundary excitations in the Quantum Hall states and in the chiral
spin states", \prb , {\bf43}, 11025-11036, (1991)
 \item{\wigner .} E.~P.~Wigner, G\"ottinger Nachr. {\bf 31}, 546,
(1932),  and also {\it Group Theory}, Academic Press, N.Y., (1959).
\item{\wilczek .} F.~Wilczek, {\it Fractional Statistics and Anyon
Superconductivity}, World Scientific, (1990).
\item {\xia .}  J.~Xia, ``Geometric invariants of the quantum Hall
effect'',  \cmp \  ,{\bf119}, 29--50, (1988).
\item{\zak .} J.~Zak, ``Magnetic translation group'', Phys.\ Rev.\ ,
{\bf 134}, A1602--1607, (1964) and ``Magnetic translation group
II: Irreducible representations'' 1607--16011, (1964);
and in  {\it Solid State Physics}  {\bf 27}, F.~Seitz, D.~Turnbull and
H.~Ehrenreich, {\bf 59}, Academic Press, (1972).
\enddocument